\documentstyle[12pt]{article}
\title{Relativistic Corrections  in a  Three-Boson System  \\ of Equal Masses}
\author{Ph. Droz-Vincent\\[2mm] LUTH  \      Observatoire de Meudon\\
5 place Jules Janssen 92195 Meudon, France}

               \date{\        }

\newcommand  {\eeq}{\end{equation}}

\newcommand  {\beq}{\begin{equation} }
\newcommand  \half {  {1 \over 2} }
\newcommand  {\ytil}{\widetilde y}
\newcommand{\yhat}{\widehat y}
\newcommand {\ztil}{\widetilde z}
\newcommand{\zhat}{\widehat z}
\newcommand  {\noi}{\noindent}
\newcommand  {\disp}{\displaystyle}
\newcommand  {\zer}{ {(0)} }

\newcommand {\alp}{\alpha}
\newcommand {\lam}{\lambda}
\newcommand{\Gam}{\Gamma}

\newcommand{\soulu}{\underline u}
\newcommand{\soulv}{\underline v}
\newcommand{\soulV}{\underline V}

\newcommand{\soulxi}{\underline \xi}
\newcommand{\soulXi}{\underline \Xi}
\newcommand{\souleta}{\underline \eta}

\newcommand{\batonR}{{\mbox{I\hspace*{-1mm}R}}}
\newcommand{\bfy}{ {\bf y} }
\begin{document}
\maketitle
\abstract{Three-body systems of scalar bosons are covariantly described in
the framework of relativistic constraint dynamics.
 With help of a  change of variables followed by a change of
wave function,   two  redundant degrees of freedom  get eliminated and the
 mass-shell constraints can be reduced to a three-dimensional eigenvalue
problem.
In general, the reduced equation  obtained by this procedure
 involves  the spectral parameter in a nonconventional manner,
but for three equal masses a drastic simplification arises at the first
 post-Galilean order: the reduced wave equation becomes a conventional
 eigenvalue problem that we treat perturbatively, computing a first order
correction beyond  the nonrelativistic limit.
The harmonic interaction is displayed as a toy model.

\medskip

\section{Introduction, Basic Equations}
\noi   A relativistic system of mutually interacting   particles can be
 described, in a manifestly covariant way, by  mass-shell constraints.
These constraints determine the evolution of a wave function which depends on
four-dimensional arguments [1-3].
 The price paid for covariance is the presence of
redundant degrees of freedom, just like in the Bethe-Salpeter (BS) approach.

\noi For two-body systems, the extra degree of freedom is trivially
 factorized out. Moreover, in that  case, the contact  with  the
 BS equation was established~\cite{sazBS} .

\noi In contrast, for three or more particles it is difficult to find an
 interaction term such that the   mass-shell constraints are compatible among
themselves,  respect Poincar\'e invariance, reproduces free-body motion
when this term vanishes,   and allow for eliminating the  redundant degrees of
freedom.

In this paper we focus on the case of three spinless particles. We thus
 consider three Klein-Gordon equations coupled by a mutual  interaction  which
should  be either  derived from the underlaying field theory (QCD for instance)
or motivated by phenomenological considerations.
 Our basic equations are
\beq    2K_a \   \Phi  \equiv
  ( p_a ^2 + 2W ) \   \Phi =  m_a^2 c^2 \    \Phi    ,    \qquad
\qquad  a,b = 1,2,3          \label{basic}     \eeq
for a wave function with three four-dimensional arguments (either  $q_a$
or $p_b$ according to the representation used).
 The relativistic "potential" $W$ must be Poincar\'e invariant and chosen
 such that the equations above are mutually compatible.


 \noi {\sl  Remark}

 \noi In principle it seems that one could also consider more
general equations involving three distinct relativistic
"potentials" $W_1, W_2, W_3$.
But, even if we leave aside the problem of mutual compatibility
which would become more complicated, in this  more general case
there is no evidence that the superfluous degrees of freedom could
be eliminated at all. It is therefore natural to focus on the
simple class of models characterized by  $W_a = W$.
 This choice is reminiscent of what is currently done in
the  two-body case, where  using the same interaction function in
both wave equations is general enough to accommodate most
realistic situations\cite{sazqk}\cite{jallou97}.

\noi For three particles, the assumption of a single interaction
function in eqs (\ref{basic}) will be justified {\em a posteriori}
by its efficiency in the task of reducing the degrees of freedom.
Of course in a future work, a justification by a contact with
field theory would be desirable.

\noi It should be emphasized that having $W_a = W$ by no means
forbids to take into account differences in the couplings that
concern each particle. In most systems of practical interest $W$
is a sum of three terms; each one of these three terms, although
it is not strictly binary (due to the three-particle forces
automatically included) still carries some two-body input (the
three-body forces being of higher order).

This will be seen for instance in the covariant harmonic potential
of Section 4, where the potential given by equation
(\ref{supplem}) includes three distinct spring constants
permitting  to implement a different interaction law inside each
 cluster; for electromagnetic interactions different
charges could be handled in a similar way.


\bigskip
 \noi Poincar\'e algebra is realized in the
same manner as for the free-particle case, say
 $P = p_1 + p_2 + p_3$ and $M= \sum  q_a \wedge p_a$.
\noi    It  is convenient to introduce relative variables with the
"heliocentric" notation:       relative-particle indices are   $A, B = 2, 3 $.
We define   the four-vectors
\beq z_A = q_1 - q_A  ,  \qquad \qquad
 y_B =  {P \over 3} -  p _B        \label{relative}   \eeq
Their  transverse parts  are  $\ztil _A , \ytil _B$. Tilde denotes the
projection orthogonal to P, for instance
$\disp  \ztil_A  =  z_A -  (z _A  \cdot P ) P / P^2 $, etc.

\medskip
\noi  With help of the identity
\beq  3 \sum p^2  \equiv  P^2 + D + 6 P^2 \Xi   \label{newnewiden}       \eeq
where
\begin{equation}
 D = 6 (\ytil  _2 ^2  + \ytil _3 ^2  +  \ytil _2 \cdot  \ytil _3 )
                                               \label{defD}  \end{equation}
  \begin{equation} \Xi =
 (P^2) ^{-2} [ ( y_2 \cdot  P )^2  +  (y _3 \cdot P ) ^2  +
 (y _2 \cdot  P )( y_3 \cdot P )  ]       \label{dejaXi}  \end{equation}
the sum of equations (\ref{basic})  yields a dynamical equation  involving the
potential.
On the other hand,  the differences of equations     (\ref{basic})
take on the purely kinematic form
\beq  (p_1 - p_A )(p_1 + p_A ) \      \Phi  =
 2   \nu _A  c^2  \      \Phi              \label{newdif} \eeq
where the half squared-mass differences are
              $\    \disp    \nu _A = \half  (m_1 ^2 - m_A ^2 )   \    $.
For the sake of  compatibility we require that $W$ commutes with  both
products $ (p_1 - p_A )(p_1 + p_A )$.

\medskip
\noi In order to achieve the  elimination of  two degrees of
freedom, we have proposed   \cite{droz3}
 a quadratic change of variables in momentum space, say
    $ p_a    \mapsto  p' _ b $ or equivalently
$$  P, y_A    \mapsto    P,  y'_A           $$
 This  transformation  can be
 characterized as a redefinition of the relative energies
   such that
\beq     (p_1 + p_A)\cdot (p_1 - p_A) = P \cdot  (p'_1 -p'_A)
                                                   \label{transfo}     \eeq
and by the requirement that it leaves $P, \ytil _2 $  and $\ytil _3$
unchanged, that is
\beq   P' = P,  \qquad  \qquad
        \ytil '_A =  \ytil _A   \label{transfobis}     \eeq
Clearly equation (\ref{transfo})  determines in closed form the longitudinal
 pieces of
$p'_2 , p'_3$ (resp. $y'_2, y'_3$) in terms of all the primitive variables
$p_a$ (resp. $P, y_A $)~\cite{saz3}.

\noi Of course, we define the new relative momenta $y'_A$
as  linearly related with the $p' _a$'s through a
formula similar to  (\ref{relative}), namely
$$ y'_A =   {P \over 3}  - p'_A                                $$
  Note that our transformation preserves Poincar\'e invariance; as a result
the generators of spacetime displacements have the usual form also in terms of
$p'_a$.

\noi        Naturally, this procedure   gives  rise to  new configuration
        variables   $z'_A$.   In general, the new variables  $z'_A , y'_B $ are
 referred to as {\sl reducible}.

\bigskip
\noi It is noteworthy that instead of $\Phi$,
we can   equivalently use a new wave function
 \beq  \Psi =
|J|^{1/2}  \Phi ,  \qquad \qquad  \
         J =
 {D (p_1, p_2 ,p_3) \over  D(p'_1 , p'_2 , p' _3 ) }       \eeq
Accordingly, the  operators $K$ and $W$ are mapped to
 $H$ and $V$ respectively,
\beq H =   |J|^{1/2}  K   |J|^{- 1/2} ,    \qquad  \qquad
 V  =   |J|^{1/2}   W      |J|^{- 1/2}       \label{map}   \eeq

\noi
In contrast to  $z_2 , z_3$, the operators
 $z'_A$ are  "formally hermitian" ({\em i.e.} symmetric) with respect
to the  Hilbert space $ L^2 (\batonR ^{12}, d^{12} p')$.
  In other words, setting

$  d^{12} p = d^4 p_1 \    d^4 p_2 \    d^4 p_3 , \qquad \qquad
  d^{12}p' = d^4 p'_1 \   d^4 p'_2 \    d^4 p'_3 ,  \qquad $   we have
 \beq   \sum (z' _A \Upsilon )^*  \Omega  \    d^{12}p' =
        \sum   \Upsilon ^*  z'_A  \Omega  \    d^{12} p'
 \eeq
whenever $\Upsilon$ and $\Omega$ are square integrable in terms of
the volume element $d^{12} p'$. Owing to (\ref{map}), $H$ and $V$
have the same property~\cite{offshell}.


\bigskip
\noi  The results of Ref.~\cite{droz3} are as follows:

a) Provided the three masses are not too much different one from
another, equations (\ref{transfo})(\ref{transfobis}) can be
inverted in closed analytic form.

\noi
Indeed our model is reliable insofar as, in the no-interaction
limit obtained by putting the potential equal to zero, one
recovers the free motion of independent particles.
The discussion of this point in Ref.~\cite{droz3} led to  require
\beq |\nu_2 + \nu_3 | <  {1 \over 24 } \sum m_a ^ 2 ,
 \qquad \quad |\nu_2 -\nu_3 | <  {1 \over 8} \sum  m_a ^2
\eeq
For instance,  if only two masses are equal, say $m_2 = m_3$,
their value is allowed to deviate from  $m_1 $ by an amount of
almost $6 \% $.


 b) Conditions (\ref{transfo}) amount to redefine  relative
 energies, in such a way that the {\em new} relative energies can be eliminated.

c)  The compatibility conditions
can be satisfied easily  in terms of our  {\em new} variables.
Actually, in view of the  compatibility requirement, a  closed form
 of the interaction  is available only in terms of the {\em new variables}.

 A typical example of a potential satisfying compatibility and
Poincar\'e invariance would be of the form
 \beq
V =f ( (\ztil '_2)^2 , (\ztil ' _3)^2 , \ztil '_2 \cdot  \ztil '_3
, P^2)
 \label{generalV}  \eeq
since  any function of $\ztil '_B  , \ytil '_C $ and $P$ commutes
with $y' _A \cdot P$.  In this equation all the arguments of $f$
are mutually commuting~\cite{admissV}.

\noi  This situation is in favor of  using $\Psi$ and the new
variables, as we shall do hereafter.

\medskip
\noi
By our transformations,
the difference equations    (\ref{newdif})  become
\beq  y'_A \cdot P \     \Psi  =
 ( { 4\over3 } \nu _A  -  {2\over 3 }   \nu_B )  c^2 \     \Psi ,
       \qquad  \qquad  {\scriptstyle  A \not= B}  \label{diffeqs}      \eeq
and the  dynamical equation of motion
 is mapped to
\beq  (3 \sum m^2  c^2  - P^2) \    \Psi=
(D + 6 P^2   \Xi + 18 V ) \    \Psi     \label{weqsumbis}          \eeq
In order to handle this equation       
  we need to express  $\Xi$
 in terms of the new variables.

\noi  Lenghty but elementary manipulations reported in  \cite{droz3}
show that
\beq    \Xi = \xi ^2 + \eta ^2 + \eta \xi      \label{defXi} \eeq
where $\xi, \eta$ are determined by the system
\begin{equation}  {2 \over 3} \xi + {1 \over 3} \eta + { \xi \eta }  +
{\eta ^2  \over 2}   =  u                \label{linxi} \end{equation}
\begin{equation}     {2 \over 3} \eta + {1 \over 3} \xi + { \xi \eta }  +
{\xi ^2 \over 2 }  =     v                 \label{lineta}    \end{equation}
 $u, v$  being   determined  as follows
   \beq   P^2 u=  y'_2 \cdot  P  + \half  y'_3  \cdot P  -
(\ytil _2 \cdot \ytil _3  + \half {\ytil} _3 ^2 )
                                                 \label{defu}  \eeq
\beq P^2 v=       y'_3 \cdot  P  + \half  y'_2  \cdot P
    -  ( \ytil _2 \cdot \ytil _3  + \half  {\ytil} _2 ^2)
                                                 \label{defv}     \eeq

\section{Three-dimensional reduction}

\medskip
\noi Now  the dependence of $\Psi $ on the {\em new} relative energies is
easily  factorized out, provided  we assume a sharp linear momentum, say
\beq  P^\alp  \Psi  =  k^\alp \Psi , \qquad \qquad   k^2 = M^2  c^2      \eeq
for some constant timelike vector $k$.
Let  $\     \widehat {   } \   $ denote the projection orthogonal to $k$. For
instance  the transverse piece of $z$ with respect to $k$ is
$ \disp \zhat = z - {z \cdot k  \over k^2} k $, etc.
 In the rest frame we have
$         {\yhat }^2 _A =  - {\bf y}^2 _A  ,
 \quad    {\zhat }^2 _A =  - {\bf z}^2 _A  $, etc.

\medskip
\noi
We make this convention that, in any  operator $F$
 depending on the dynamical variables, the underline indicates that we replace
$ y'_A \cdot P$ by
 $ \disp   ( {4\over 3} \nu _A  -  {2 \over 3} \nu _B ) c^2$
and  $P^\alp$ by $k^\alp$,  hence  $P^2$ by $M^2 c^2$. Let us write   this
symbolically
$$ {\underline F } = {\rm subs.}
 (  y'_A \cdot P = ( {4\over 3} \nu _A  -  {2 \over 3} \nu _B ) c^2 ,
    \qquad  P^\alp = k^\alp , \qquad  F )                      $$
It is clear that  $F$ reduces to $\underline F$  on the "mass-momentum shell"
(defined as the subspace of solutions to the   mass-shell constraints which
are eigenstates of total linear momentum).

\noi  Note that $\soulXi$   depends only on  $\yhat_2 , \yhat_3$.

\noi   Equation     (\ref{weqsumbis}) yields the {\em  reduced equation}
\beq (3 \sum m^2 - M^2 ) c^2   \psi =
 (  6 ({\yhat}^2 _2 + {\yhat}^2 _3 + \yhat _2 \cdot \yhat _3 )
+ 18 \soulV +  6  M^2 c^2 {\underline \Xi}    )\   \psi
                                                    \label{redeq3}      \eeq
for a reduced wave function $\psi$ which depends on three-dimensional
 arguments only
(say  $\yhat _2 , \yhat _3$ in the momentum representation).

\section{Equal Masses}

\noi   Fortunately, { \em in the case of three equal masses}, say
$m_a = m$
 we have this further simplification that
 $\nu_A = 0$, which finally renders $\soulu , \soulv $ of the order of
$1/c^2$.
More precisely   (\ref{defu})(\ref{defv})   entail
\beq  M^2 c^2 \soulu = - (\yhat _2 \cdot \yhat_3  + \half \yhat _3 ^2 )
                                   \label{soulu}                          \eeq
\beq  M^2 c^2 \soulv = - (\yhat _2 \cdot \yhat_3  + \half \yhat _2 ^2 )
                                    \label{soulv}                         \eeq
hence  $\soulu$ and $\soulv $ to be inserted into  the reduced version of the
system   (\ref{linxi})(\ref{lineta}).
Solving for ${\underline \xi} , {  \underline \eta}  $   we obtain
\beq  \soulxi = 2 \soulu  - \soulv      + O (1/c^4)                     \eeq
\beq  \souleta = 2 \soulv  - \soulu     + O (1/c^4)                     \eeq
Inserting this into (\ref{defXi}) we get
\beq  \soulXi = 3 ( \soulu ^2 + \soulv ^2 - \soulu  \cdot \soulv )
   +  O ( 1/c^6)          \label{apprXi}               \eeq
correcting a misprint in  the higher-order term of equation (100)
of  Ref.~\cite{droz3}.
Since the leading term in  $\soulXi$ is $O(1/c^4 )$,
we have   defined   $\Gam$  by setting
\beq     M^4 c^4 \soulXi  = \Gam = \Gam _{(0)} +
  {1 \over c^2}   \Gam _ {(1)} +  \cdots             \label{defGam}     \eeq
where            $\Gam _\zer ,  \Gam _ {(1)}, \cdots$
 remain finite when $c \rightarrow \infty$.
In view of  (\ref{apprXi}) it is clear that
$$      M^4 c^4 \soulXi =
3 M^4 c^4 ( \soulu ^2 + \soulv ^2 - \soulu  \cdot   \soulv )
        +    O (1/c^ 2 )            $$
 We compute  respectively    $ \soulu ^2 ,  \soulv ^2$  and
  Hence
\beq     \Gam _{(0)} =  {3\over 4}   \      \{
({\yhat _2}^2 )^2 +  ({\yhat _3}^2 )^2 + 4   (\yhat _2 \cdot \yhat _3 )^2
+ 2  ({ \yhat _2}^2  +  { \yhat _3}^2 ) \      ( \yhat _ 2 \cdot \yhat _3)
 - {\yhat _2 }^2  {\yhat _ 3 }^2    \}
                           \label{Gamzero}                    \eeq
expression valid only for  three equal masses (this formula was given in
Ref. \cite{droz3} without proof).
Note that  $\Gam$ is a positive operator and would survive  in the absence of
interaction.

\medskip
\noi
For three equal masses, equation (\ref{redeq3}) takes on the form
\beq    (9 m^2 - M^2 ) c^2   \psi =
6  ( {{\yhat} _2} ^2   +  {\yhat} _3 ^2   + \yhat _2 \cdot \yhat _3 ) \   \psi
+  18  \soulV  \psi   +   {6 \over M^2  c^2} \    \Gam \     \psi        \eeq
Defining
\beq    6  \lam    =  (M^2   - 9 m^2 ) c^2       \label{deflam}     \eeq
and using the rest frame
 (where
 $  \yhat _A \cdot  \yhat _ B =  -  \bfy _A \cdot  \bfy _B  $)  we can write
\beq  \lam  \psi =
 (   {\bf y} _2  ^2  + {\bf y} _3 ^2  + {\bf y _2} \cdot {\bf y} _3   ) \
 \psi   -  3 \soulV   \     \psi   -  {\Gam  \over  M^2  c^ 2}   \     \psi
                                        \label{redeqeqmass}          \eeq
\noi Naturally $\Gam_{(0)}$ admits an expression identical to
     (\ref{Gamzero})    in terms of  $\bfy _2  ,  \   \bfy_3$.

\noi
In spite of being three-dimensional, the reduced  equation
 (\ref{redeqeqmass})
 as it stands, is  more problematic than an ordinary eigenvalue problem.
Even if the interaction  doesnot depend on the total energy (that is: $V$
doesnot depend on $P^2$)   the term
$P^2 \Xi$  in (\ref{weqsumbis}), which  has no counterpart in two-body systems
and yields   $M^2  c^2 \soulXi$  in
(\ref{redeq3}), brings out some energy dependence.
 It follows that
  (\ref{redeqeqmass}), is  not
 a conventional eigenvalue equation: through (\ref{deflam})
 the operator to be diagonalized   depends on its own eigenvalue.
 This complication  is by no
means a drawback special to our model. As emphasized in     \cite{todsazriz}
it plagues  most   relativistic wave equations; the mathematical  theory
of this situation is  rather involved, but   fortunately this difficulty can
 be  more easily handled in a perturbation scheme,  {\em provided the
 unperturbed equation is not energy dependent}.

\noi In the rest of this paper we focus on the first relativistic corrections.
Therefore we  solve  (\ref{redeqeqmass})  after  expansion  in powers
 of  $1 / c^2$, taking  (\ref{deflam}) into account, say
$  \disp     M^2 =  9 m^2   +  6  {\lam  /  c^2 }$.
In principle, the exact analytic expression for $\Gam$
is known, and is itself a series in  $1 / c^2$. In fact the knowledge of
$\Gam _\zer$ is sufficient for our purpose.
Assuming that  $\lam$  remains finite in the nonrelativistic limit,  we select
these solutions that are in some sense "close to" the nonrelativistic
Schroedinger equation obtained by dropping
 $ 1 /  c^2$  in   (\ref{redeqeqmass}). 

\noi This development is justified insofar as the velocity of light can be
considered as large with respect to some velocity formed with help of the
physical parameters defining the system. Practically, the constituent masses
and the coupling constants involved in the interaction term must be combined
as to form a quantity having the dimension of speed. In principle one should
check that this "characteristic velocity" actually has something to do with
the average velocities of the constituent particles in the slow motion
approximation.

\noi The legitimity and the limitations of this procedure vary according to
the analytic shape of the interaction  term and must be discussed in each
specific case.

\subsection{Post-Galilean Approximation}

\medskip
\noi  Let us start expanding in powers of $\disp 1/c^2$. Using the rest frame,
and assuming that
\beq    \psi = \psi _\zer +   {1 \over c^2}  \psi _{(1)} +   \cdots ,
   \qquad  \         \soulV =  \soulV _ \zer
   + {1 \over c^2} \soulV _{(1)} + \cdots    \label{devsoulV}      \eeq
the {\sl zeroth order} approximation   to  (\ref{redeqeqmass}) yields   the
nonrelativistic limit
\beq  \lam _0   \psi _\zer  -
(   {\bf y} _2 ^2   +  {\bf y} _3 ^2   + {\bf y} _2 \cdot {\bf y} _3    )
\      \psi _\zer    + 3 \soulV _\zer \      \psi _\zer
                  = 0                  \label{0rest}          \eeq
Setting
  \beq       E_ \zer 
 =   {\lam _\zer  \over m},   \qquad \qquad
   U = -{3 \over m} \   \soulV _\zer         \label{defEnr}  \eeq
equation (\ref{0rest})  can be re-written as
\beq      E_ \zer     
  \     \psi _\zer  =    {1\over m}
  ( {\bf y} _2 ^2   +  {\bf y} _3 ^2   + {\bf y} _2 \cdot {\bf y} _3    )
 \      \psi _\zer    +  U  \psi _\zer            \label{nrlim}         \eeq
which is  similar to the  Schroedinger equation of a
nonrelativistic problem with three equal masses ({\em except perhaps} for
 complications resulting from
a possible dependence of $V$ on $P^2$).
Indeed  we consider equal masses, thus $m = 2m_0$ where $m_0$ is
the reduced mass of either of particles 2, 3, with respect to particle 1.
The first operator in the r.h.side is nothing but the kinetic
energy  for a nonrelativistic system of three masses $m$, when  the
center-of-mass motion has been separated.

\medskip
\noi  At the {\sl  first order} in $1/c^2$  we can, in the last term of
equation (\ref{redeqeqmass}),  replace $\Gam$ which depends on $M^2$,
 by $\Gam _{(0)}$, which doesnot. In view of  (\ref{deflam}), in this last
 term,  we can also replace  $M^2$ by $9m^2$. Hence
\beq    \lam   \psi  =
   ( {\bf y} ^2 _2 + {\bf y} ^2 _3 + {\bf y } _2 \cdot {\bf y  _3}
  -  3\soulV
 -  {\Gam_{(0)}   \over  9m^2  c^2} )  \psi     \label{order1rest}     \eeq
with $\Gam _ {(0)}$  bi-quadratic in $\bf y$.
Inasmuch as $V$ is not energy-dependent,  the above equation still has the
structure of a nonrelativistic eigenvalue problem, and can  be solved
by treating   the last term    as a perturbation.

\noi More care is needed for  most  realistic  potentials,  for which
  $V$ depends on  $P^2$, hence $\soulV$
depends on $M^2 c^2$.
Fortunately, in several cases, this  dependence is  of
higher order, so that it can be accounted for by addition  of  an extra
 perturbation  term,  as follows.
Assuming that  $\soulV$ is as in
       (\ref{devsoulV})  we have     
\beq    \lam  \psi  =
  ( {\bf y} _2 ^2   +  {\bf y} _3 ^2   + {\bf y} _2 \cdot {\bf y} _3    )
\psi   - 3  \soulV _\zer  \psi
- {1\over c^2}  (  {\Gam _ \zer  \over 9m^2}  +  3 \soulV _{(1)}  )
\      \psi                  \label{order1bis}       \eeq
Since we do not go beyond first order, let us write
$ \disp     \lam  =  \lam _\zer  +  {1\over c^2}   \lam _{(1)}    $.

\noi
For any {\em  nondegenerate} level $\lam$,  we have
\beq   \lam _ {(1)}    =     - \   < {\Gam _ \zer \over  9 m^2 }      +
 3  \soulV _ {(1)}      >                             \label{correc}      \eeq
where the expectation value  must be  calculated in the unperturbed
 eigenstate  $\psi _\zer$.

\bigskip
\noi    {\bf Binding energy.}

\noi Now we are in a position to calculate, at first post-Galilean order,
  the   binding energy of a bound state.
This quantity is usually defined through the (linear) mass defect~\cite{ouch}.
 So let us evaluate     $M -  \sum m  =  M - 3 m  $.
Taylor expansion of  (\ref{deflam})  yields
\beq  (M - 3 m)c^2   =
   {\lam \over  m }  -
 {\lam ^2 \over 6 m^3 c^2}   +   O(1/c^4 )     \label{bind3}  \eeq
\beq  (M - 3 m)c^2   =
   {\lam _ \zer \over  m }   + {1 \over c^2}   \
        ( {\lam _ {(1)}  \over m}   -   {\lam_ \zer ^2  \over 6 m ^3 } )
                     + O(1/c^4 )                      \label{correcbind}  \eeq
which yields the first  correction to binding energy.

\subsection{Jacobi's coordinates}

Equation       (\ref{order1bis})   amounts to
 a nonrelativistic problem, formulated in terms of the canonically conjugate
 variables  ${\bf z}'_A ,  \bfy '_B$.
Before we turn to the harmonic interaction it is convenient to introduce
  {\sl Jacobi's coordinates} that have the virtue of simplifying the
expression of the kinetic energy.    So we perform a {\sl linear change} from
$ {\bf z}' _A , {\bf y} ' _B $ to $ {\bf R} _A , {\bf \Pi }_B $, as follows.

\noi  For {\sl three equal masses},
the Jacobi coordinates $ {\bf R}_2  ,   {\bf R}_3  $         associated with
 ${\bf q} '_2 ,     {\bf q} ' _3$, are defined by the formulas \cite{rich}
\beq  {\bf R}_2 =  {\bf q} '_2  -  {\bf q} '_3  ,   \qquad    \qquad
  {\bf R}_3 =   {1 \over \sqrt{3} }\
     (2    {\bf q} '_1  -   {\bf q} '_2  - {\bf q} '_3   )             \eeq
in other words
\beq  {\bf R}_2 =  - {\bf z } ' _2  + {\bf z} '  _3  ,  \qquad  \qquad   \
 {\bf R }_3 =   {1 \over \sqrt{3} } \    ( {\bf z } ' _2  + {\bf z} '  _3 )
                                         \label{jacobi}                \eeq

\noi
Inverting (\ref{jacobi}) yields
\beq    {\bf z } ' _2 = \half (\sqrt{3}  {\bf R}_3 -      {\bf R}_2 )  \qquad
\qquad  \     {\bf z } ' _3 =  \half (\sqrt{3}  {\bf R}_3 +   {\bf R}_2 )
                                     \label{InverJac}               \eeq
Since (\ref{jacobi}) is   a  linear  transformation,
it is easy to determine conjugate momenta, say
$ {\bf \Pi}_2 , \       {\bf \Pi} _3  $, such that
$  [{\bf R}_2  , {\bf \Pi} _2 ] =
    [{\bf R}_3 , {\bf \Pi} _3 ]= i \delta $  and
$  [{\bf R}_2  , \bf \Pi _3 ] =    [{\bf R}_3 , {\bf \Pi} _2 ]= 0$,  etc.
We find
\beq  {\bf \Pi} _2 =  - \half     {\bf y} _2 + \half  {\bf y} _3 , \qquad \qquad
\    {\bf \Pi} _3 =  { \sqrt{3}   \over 2}  \  ({\bf y} _2  +{\bf y} _3 )
                                               \label{jacobiconj}     \eeq
Hence inverse formulaes
\beq  {\bf y} _2 = - {\bf \Pi} _2 +  {1 \over \sqrt{3} } \    {\bf \Pi} _3 ,
 \qquad  \qquad   \        \
      {\bf y} _3 =  {\bf \Pi} _2 +  {1 \over \sqrt{3} }  \    {\bf \Pi} _3
                                             \label{invjaconj}   \eeq

\medskip
\noi
   In equation     (\ref{nrlim})
 kinetic energy was expressed in terms of the heliocentric coordinates.
But with help of (\ref{invjaconj})      
 we can write
\beq   {\bf y} _2 ^2  +  {\bf y} _3 ^2  +   {\bf y} _2 \cdot  {\bf y} _3  =
    {\bf \Pi} _2  ^2       +   {\bf \Pi} _3  ^2     \label{kiny}    \eeq
Now (\ref{nrlim}) may be re-written in terms of the Jacobi coordinates.
For the {\em total} kinetic energy we  have
\beq  \sum  {  {\bf  p} ^2   \over 2m }   =
{ {\bf  P}^2   \over  6 m}
  +  {1 \over m} \    (  {\bf \Pi} _2 ^2       +   {\bf \Pi} _3 ^2  )
                                             \label{IV}     \eeq

\noi
In order to compute the first relativistic corrections we need to evaluate
also $\Gam _ \zer$ in terms of ${\bf \Pi} _2 ,  \      {\bf \Pi} _3$.
So we must insert  (\ref{invjaconj}) into  (\ref{Gamzero}).
  To this end we can write
\beq {4\over 3} \Gam_\zer = A^2 + B^2 + 4C^2  + 2 (A+B)C - AB    \eeq
with this  notation
$$ A = ({\bf y}_2)  ^2 , \qquad \qquad \  B=   ({\bf y}_3)  ^2 ,\qquad \qquad
\             C = {\bf y}_2   \cdot     {\bf y}_3               $$
$$ {\cal  A} = ({\bf \Pi }_2)  ^2 ,
\qquad \qquad \  {\cal  B}=   ({\bf \Pi}_3)  ^2 ,
\qquad \qquad \         {\cal  C} = {\bf \Pi}_2   \cdot   {\bf \Pi}_3        $$
From  (\ref{invjaconj}) we get
$$ A =    {\bf \Pi}_2 ^2 -
       {2\over \sqrt{3}}   {\bf \Pi}_2   \cdot     {\bf \Pi}_3
    + {1\over 3}  {\bf \Pi}_3 ^2       $$
$$ B =    {\bf \Pi}_2 ^2     +
       {2\over \sqrt{3}}   {\bf \Pi}_2   \cdot     {\bf \Pi}_3
    + {1\over 3}  {\bf \Pi}_3 ^2       $$
$$  C=                            {1\over 3}  {\bf \Pi}_3 ^2   -
                                  {\bf \Pi}_2 ^2                         $$
in other words
$$ A =    {\cal  A}  -  {2\over \sqrt{3}} {\cal  C}
                               + {1\over 3}  {\cal  B}                $$
$$ B =    {\cal  A}  +   {2\over \sqrt{3}} {\cal C}
                               + {1\over 3}  {\cal  B}                $$
$$  C=   {1\over 3} {\cal  B} -  {\cal  A}                             $$
Inserting into (\ref{Gamzero}) we get
$$   {4\over 3} \Gam_\zer = ({\cal  A}+{\cal  B})^2 + 4 {\cal  C}^2     $$
\beq  {4\over 3} \Gam_\zer =    ({\bf \Pi }_2  ^2 )^2   +
 ( {\bf \Pi}_3  ^2 )^2    +  2   {\bf \Pi }_2  ^2      {\bf \Pi}_3  ^2
  +  4  ( {\bf \Pi}_2   \cdot     {\bf \Pi}_3 )^2
                                   \label{jacogamzer}          \eeq

\section{The Covariant Harmonic Potential}
\noi            In order to test the formalism,
it is natural to consider first a toy
 model, namely  the harmonic oscillator.


\noi Harmonic interactions are implemented through the potential
 \beq V
=      \kappa_{12}  \   {\widetilde {(q'_1 - q'_2 )}} ^2  +
\kappa_{23} \   {\widetilde {(q'_2 - q'_3 )}} ^2    + \kappa_{13}
\    {\widetilde {(q'_1 - q'_3 )}} ^2 \label{supplem} \eeq
 where
$\kappa _{ab} $  are positive coupling constants.


\noi If, for the sake of simplicity,  we assume that all these
constants  are equal we obtain this version
  \beq       V = \kappa  \sum _{a<b}
{{\widetilde {(q'_a -q'_b)} } }^2   =
         2  \kappa     \{  ({\ztil}'_2  ) ^2  +   ( {\ztil}'_3  )^2  -
       {\ztil}'_2   \cdot    {\ztil}'_3      \}    \label{oscar}     \eeq
where  $\kappa$ is a positive constant.   The identity
\beq     \sum   _{a<b}     {{\widetilde { (q'_a -q'_b)} } }^2  \equiv
2  \{  ({\ztil}'_2) ^2  +   { (\ztil}'_3 ) ^2  -
       {\ztil}'_2   \cdot    {\ztil}'_3     \}
                                            \label{oscarid}  \eeq
reads, after reduction to the rest frame
\beq      \sum  ( {\bf q}'_a -  {\bf q}' _b) ^2 =
  2  \      [  {\bf z'} _2 ^2 + {\bf z'} _3 ^2  -
            {\bf z'} _2 \cdot  {\bf z'} _3   ]       =
  {\bf z'} _2 ^2 + {\bf z'} _3 ^2 + ( {\bf z'} _2 - {\bf z'} _3 )^2
                                            \label{V}           \eeq
With help of (\ref{InverJac})
  we have
   ${\bf z'} _2 - {\bf z'} _3 = - {\bf R}_2 $ and
$${\bf z'} _2 ^ 2  =   {1 \over 4} \
  (3 {\bf R}_3 ^2 - 2 \sqrt{3}   {\bf R}_3 \cdot {\bf R}_2 +
{\bf R}_2 ^2   )     $$
$${\bf z'}_3 ^ 2  =   {1 \over 4} \
  (3 {\bf R}_3 ^2 + 2 \sqrt{3}   {\bf R}_3 \cdot {\bf R}_2
   + {\bf R}_2 ^2   )     $$
Note that
$$      2  \      [  {\bf z'} _2 ^2 + {\bf z'} _3 ^2  -
                 {\bf z'} _2 \cdot  {\bf z'} _3   ] =
{3 \over 2}
({\bf R}_2 ^2  +  {\bf R}_3 ^2 )                          $$
for all choice of units. Hence we obtain
\beq   \sum ({\bf q'}_a - {\bf q'}_b)^2   =  {3 \over 2}
({\bf R}_2 ^2  +  {\bf R}_3 ^2 )                      \label{VI}     \eeq
which exhibits the $O_6$ invariance of our potential $U$.

\noi   Finally equation   (\ref{order1rest}) takes on the form
\beq  \lam  \psi  =   (  {\bf \Pi}_2 ^2  +   {\bf \Pi}_3 ^2 )  \psi
    +   {9 \over 2}\kappa \    ( {\bf R}_2 ^2  +   {\bf R}_3 ^2 )       \psi
      -  ({\Gam _\zer  \over  9m^2 c^2})            \psi
                                             \label{su6}               \eeq

For the moment, let us consider  the zeroth-order approximation and divide
 by $m$.   We obtain the
Schroedinger equation of a non-relativistic three-body oscillator with
equal masses, written in Jacobi coordinates
$  {\bf R}_2 ,    {\bf R}_3 ,   {\bf \Pi}_2  ,  {\bf \Pi}_3  $
(the $SU_6$ invariance of the nonrelativistic limit would become manifest if we
were to choose an appropriate unit of lenght).

\noi In order to make the contact with textbook notations
\cite{rich},  we may define
$$    K =    { 6  \kappa  \over  m}  $$
The nonrelativistic potential is
$$ U = -{3 \over m} \soulV  =
 {K \over 2}            \sum ( {\bf q'}_a - {\bf q'}_b  )^2     $$
At the zeroth order  the ground-state wave function is  a Gaussian,
 as well in the coordinate as in the momentum representation. We have better
to choose the latter,  where the operator $\Gam$ is multiplicative. Then  the
unperturbed ground state is
\beq    \psi _\zer =   \phi   =
{\rm const.} \      \exp \{ -  {1 \over  3  \sqrt {2 \kappa}}
  ( {\bf \Pi}_2 ^2 +    {\bf \Pi}_3 ^2    )  \}                      \eeq

\medskip
\noi
In order to check the validity of  expanding in powers of $1/c$ we observe
that the quantity      $\disp  {1 \over m} \sqrt{ 3K /m}$
has the dimension of a {\em squared} velocity.
The velocity obtained by taking its
square root is a characteristic of the system  and should be  reasonably
small  with respect to the speed of light.

\noi Since $V$ doesnot depend on $P^2$ it follows that $\soulV$ doesnot
depend on $M^2$, thus equation  (\ref{nrlim}) is an
 eigenvalue problem  in the conventional  sense.

\noi With notations  (\ref{defEnr}) we have for  the $n^{\rm th}$ level
of the unperturbed harmonic oscillator,
\beq     E_ \zer
  =  \sqrt{ 3K /m}\    (3 + n)   ,
 \qquad \qquad  n= 0,1,2 \cdots  \infty           \eeq
It is convenient to set
\beq             \omega =    \sqrt{ 3K /m}
=   {3 \over m} \     \sqrt{2 \kappa}      \label{defomega}             \eeq
Indeed, for the  ground state  we have in particular
\beq   E_\zer =  3 \omega , \qquad  \
    \lam_\zer = 9 \sqrt{2 \kappa}   = 3m \omega     \label{groundst}    \eeq

\bigskip
\noi    Let us now consider the first post-Galilean contribution;
first order perturbation theory applies as usual.

We focus on the ground state;
in order to compute the first  correction,
 we need  to evaluate the expectation value of $\Gam_\zer$ in the  state
   $ \psi _ \zer $.
At this stage    we observe  that $\Gam _ \zer$
 is a homogeneous function of fourth degree in the
 six-dimensional vector     $ X  =  ( {\bf \Pi}_2  ,  {\bf \Pi}_3 ) $.
It follows that, with obvious notations, $\alp$ being any  constant
\beq    <\Gam _\zer  >   =        \alp ^{-4}    \       \int
 {\rm e} ^ { -  X^2 } \     \Gam _\zer  (X) \       d^6 X
                                       \label{Gamzerunit}             \eeq
provided that
 $  \disp   \alp ^2  =    {2 \over 3}   ( {2 \kappa}) ^ {- 1/2}       $,
which corresponds to
$ \disp  \psi _ \zer = {\rm const. e^ {- \alp ^2  X^2 /2} }   $.
Therefore  it is sufficient to carry out the calculation in the case where
$\alp =1$, so let us provisionally choose the unit of lenght such that
$\kappa = 2 / 9$.

\medskip
\noi It is convenient to note that
$ \phi =    \phi_2 \   \phi_3        $
introducing the normalized functions
$$ \phi _A = \pi ^{- 3/4}  \    \exp (- \half  {\bf \Pi}_A ^2 )   $$
As an operator ${\bf \Pi} _A$
   doesnot affect     $\phi _B$ when $B \not= A$.
Moreover  $\phi _2 , \phi _3$ are normalized to unity, so we have that
$$   < ({\bf \Pi}_A ^2 )^2 >  =
 < \phi _A ,   ({\bf \Pi}_A ^2 )^2 \    \phi _A  >                   $$
For instance we obtain
\beq  < ({\bf \Pi}_2 ^2 )^2 > =  {15 \over 4}               \eeq
and in the same way
\beq  < ({\bf \Pi}_3 ^2 )^2 > =  {15 \over 4}               \eeq

\medskip
\noi  Further we have
$$< \phi _2 \phi_3 ,\    {\bf \Pi}_2 ^2 \      {\bf \Pi}_3 ^2
 \       \phi _2 \phi_3 >
   =  < \phi _2 ,  {\bf \Pi}_2 ^2 \     \phi_2 > \
        < \phi _3 ,   {\bf \Pi}_3 ^2  \     \phi_3 >                $$
but we compute easily
$$  < \phi _2 ,   {\bf \Pi}_2 ^2 \     \phi_2 >   =
    < \phi _3 ,   {\bf \Pi}_3 ^2 \     \phi_3 >   ={3 \over 2}       $$
thus
\beq  <  {\bf \Pi} _2 ^2   {\bf \Pi} _3 ^2 >  =  {9 \over 4 }          \eeq

\medskip
\noi  Finally,  if   the $\Pi ^j _ A$ are the coordinates of the three-vector
 $ {\bf \Pi} _A$,
 we have that
\beq   ( {\bf \Pi}_2   \cdot  {\bf \Pi}_3 )^2  =
 ( \Pi _2 ^1  \Pi_3 ^1  +  \Pi _2 ^2  \Pi_3 ^2   +   \Pi _2 ^3  \Pi_3 ^3 )^2
                                                              \eeq

For the sixfold integral
     \beq             <  ( {\bf \Pi}_2   \cdot  {\bf \Pi}_3 )^2  >  =
\pi ^{-3}  \     \int    ( {\bf \Pi}_2   \cdot  {\bf \Pi}_3 )^2
                     {\rm e} ^ {-   {\bf \Pi}_2 ^2  - {\bf \Pi}_3 ^2 }
d^3    {\bf \Pi}_2   d^3     {\bf \Pi}_3                     \eeq
we find
\beq <  ( {\bf \Pi}_2   \cdot  {\bf \Pi}_3 )^2  >  = {3 \over 4}    \eeq

\noi  Linear combination of all  these results yields, according to
    (\ref{jacogamzer})
$$   <  \Gam _\zer > \  =    {45 \over 4}  =  11 + 1/4                  $$

\medskip
\noi
In view of (\ref{Gamzerunit}) we now revert to an arbitrary unit of lenght
and write
\beq           <\Gam _\zer > =   {405 \over 8}   \kappa              \eeq
Apply this result to equation (\ref{correc})  where   $\soulV _{(1)}$
is supposed to vanish,
\beq  \lam _{(1)}  =  - {45 \over 8}  {\kappa \over m^2}
                               \label{lam1}       \eeq
It is interesting to evaluate the relative importance of this correction.
For this purpose consider the quantity
$$ {\Delta \lam  \over \lam} =
                  {1 \over c^2 }  {\lam _{(1)}  \over \lam_\zer}  $$

\noi
In view of   (\ref{defomega})(\ref{groundst})(\ref{lam1}) we  finally obtain
\beq   {\Delta \lam  \over \lam} =
  - {5\over 16}  { \sqrt{2 \kappa} \over  m^2 c^2 }    =
 - {5 \over 48}  \     {\omega \over m c^2 }                              \eeq
Remind that    $  {\omega /m  } $   is the square of the
characteristic velocity.

\bigskip
\noi
Now inserting  $\lam _\zer$ and $\lam_{(1)}$ respectively given by
(\ref{groundst})  and (\ref{lam1}),    into equation (\ref{deflam}) we obtain
up to $O (1/c^6 )$
\beq   {M^2 \over m^2 } - 9 =
 {54 \sqrt{2 \kappa}  \over m^2 c^2  }  -
    {135  \kappa  \over 4 m^4 c^4 }
                                    \label{Mmkappa}           \eeq
or  equivalently
\beq    {M^2 \over m^2 } - 9 =    18  {\omega \over m c^2}
    - {15 \over 8}     ( {\omega \over m c^2} )^2
                                     \label{Mmomega}         \eeq

\noi  In principle these formulas permit to calculate $M$ at the
first post-Galilean order when  $\kappa$ and $m$ are given.

\noi   But in practice one may be
interested in a naive model of baryon. In this case it is natural to fix
  M  ( {\em e.g.} the proton
mass) and adjust $m$ and $\kappa$ in agreement with  (\ref{Mmkappa}).
The most  simple  possibility is to choose first the ratio $\disp {M / m}$
 within reasonable limits discussed below, then extract $\kappa$ from
(\ref{Mmkappa}) or alternatively extract  $\disp {\omega \over m c^2} $
  from    (\ref{Mmomega}).
      In this procedure the choice of  $M/m$
 must allow for a reasonable value of the characteristic velocity.
More precisely,  $\disp  {\omega \over m c^2} $  must be small enough
in order to justify  our  first-order treatment.
It is clear that the more   $M/m$ exceeds $3$, the more
our system is rapid.

\bigskip
\noi {\sl Example}:

 \noi If  $  \disp   {M \over m}  = 3.03 $,
solving  (\ref{Mmomega}) yields
 $$ {\omega \over m c^2 } = 0.0100     $$
so that the critical velocity is about    $10$ percent of the velocity of
light.
If  $M$ is  the proton mass  ($Mc^2 = 920$   MeV)   we  find
$ m c^2 =  303.6$     MeV for the constituent quark mass.

\section{Conclusion}

Our basic equations involve a unique interaction term and are
tailored for allowing elimination of the redundant variables
implied by manifest covariance. In most relevant cases, the
interaction term looks as if it were  made of two-body
contributions.
 In fact, the two-body nature of these contributions {\em is not}  exact,
because the transformation from original coordinates to the reducible ones
somehow mixes the individual variables.
For instance the reducible relative variable $ {\ztil}'_A$ does not exactly
 match  the cluster   (1A), and so on.
 This situation can be interpreted as due to
genuine three-body forces that we have  automatically  introduced in order to
ensure the mutual compatibility of the constraints and the possibility
 of a 3D-reduction.

\noi In general, this reduction gives rise to a non conventional
eigenvalue problem.

Starting from manifestly covariant basic equations offers several
advantages:

\noi Conceptually we realize that a general description of the
system must exist even before we assign a sharp value to $P$. And
before we impose a sharp value to $P$,  there is no rest frame
available yet ($P$ being just an operator) which seems to discard
a three-dimensional formulation {\em at this stage}.

Our approach  yields a Schroedinger-like equation only {\em at the
end}; the reduced wave equation (\ref{redeqeqmass}) contains a
post-Galilean correction which would hardly be derived from an
{\em a priori} 3D-theory.

\noi Another  motivation in favor of constraint dynamics  is  the
fact that  the contact with  quantum field theory is  more easy in
a covariant framework. Actually, this contact has been thoroughly
established in the two-body case~\cite{tod71}  where constraint
dynamics inspired an improved way of summing Feynman's diagrams
(see the "constraint diagrams" exhibited by Jallouli and
Sazdjian~\cite{jallou97}) . Of course, further work is still
needed in order to determine if the mass-shell constraints of a
three-body model also can suggest similar simplifications in the
three-body case.

\bigskip
\noi We performed  a systematic expansion in powers of $1/c^ 2$.
At least for three particles with equal masses, the
non-relativistic limit has familiar features  and the first
post-Galilean formulation is tractable:
 at this order the reduced wave equation is similar to
 a  non-relativistic equation modified by  an overall perturbation of
kinematic origin,   supplemented by an additional term which stems from the
possible energy dependence of the interaction potential.

Within this framework it is possible to compute for instance the first
 relativistic correction to  the binding energy of three given
(equal) masses bound by a given interaction.
Or alternatively, in a simple naive model like the harmonic oscillator,
one may determine the free parameters (constituent mass and/or coupling
constant) in order to fit a fixed value of the ground-state mass.

\noi Although we focused on the first post-Galilean corrections, let us
stress that our formalism is ready for use at higher orders, with
help of equation (\ref{redeqeqmass}) where $\Gam$, or equivalently
$\soulXi$, must be suitably expanded.

 \noi Future  work is needed
however, for concrete applications: we plan to  implement spin,
consider the case of unequal masses, and
 improve the  contact with other approaches~\cite{bij}.
In particular,  it may be interesting to re-visit the BS equation in terms of
the reducible variables employed here.

\medskip
\noi     Finally, our approach seems to be  more specially  designed for
 confined systems.
 When nonconfining forces are present, the occurence of scattering states
may rise the question of cluster separability which is not
addressed here. But even so, our  picture may provide, if not a
complete theory, at least  a reasonable effective {\em model}
valid  in  the  sector of   bound states.



\begin{thebibliography}{3}
\bibitem{admissV}
 A straightforward
generalization would include a dependence on the square of the
pseudo-vector $\varepsilon ^ {\alp \mu \nu  \rho }\    {\ztil
}'_{2  \mu} {\ztil '}_{3 \nu} P_\rho $.

The complete list of the dynamical variables that are translation
invariant and commute with $y'_2 \cdot P$ and  $y'_3  \cdot P$ is
given by the vectors  $\ztil '_2 ,\ztil '_3 ,   y'_2 , y'_3 , P $.
Lorentz invariance is respected when $V$ is  any  function of the
pairwise scalar products of these vectors;
but in general, going  beyond the form (\ref{generalV}) of $V$ may
require a  factor ordering of its analytic expression.
\end{thebibliography}
\end{document}